\documentclass[aps,pre,showkeys,superscriptaddress,showpacs,floatfix,twocolumn]{revtex4-1}
\usepackage{graphicx}
\usepackage[ansinew]{inputenc}
\usepackage[tbtags]{amsmath}
\usepackage{amssymb}
\usepackage{float}

\newcommand{\beq}{\begin{equation}}
\newcommand{\eeq}{\end{equation}}
\newcommand{\bc}{\begin{center}}
\newcommand{\ec}{\end{center}}

\begin{document}

\title{Impact of noise and damage on collective dynamics of scale-free neuronal networks}

\author{D.~Holstein}
\affiliation{Department of Physics $\&$ I3N, University of Aveiro, Campus Universit\'{a}rio de Santiago, 3810-193 Aveiro, Portugal}
\author{A.~V.~Goltsev}
\email{goltsev@ua.pt}
\affiliation{Department of Physics $\&$ I3N, University of Aveiro, Campus Universit\'{a}rio de Santiago, 3810-193 Aveiro, Portugal}
\affiliation{A.~F.~Ioffe Physico-Technical Institute, 194021 St.~Petersburg, Russia}
\author{J.~F.~F.~Mendes}
\affiliation{Department of Physics $\&$ I3N, University of Aveiro, Campus Universit\'{a}rio de Santiago, 3810-193 Aveiro, Portugal}

\date{\today}

\begin{abstract}
We study the role of scale-free structure and noise
in collective dynamics of neuronal networks.
For this purpose, we simulate and study analytically a
cortical circuit model with stochastic neurons.
We compare collective neuronal activity of networks with different topologies: classical random graphs and scale-free networks.
We show that, in scale-free networks with divergent second moment of degree distribution, an influence of noise on neuronal activity is strongly enhanced in comparison with networks with a finite second moment.
A very small noise level can stimulate spontaneous activity of a finite fraction of neurons and sustained network oscillations.
We demonstrate tolerance of collective dynamics of the scale-free networks to random damage in a broad range of the number of randomly removed excitatory and inhibitory neurons.
A random removal of neurons leads to gradual decrease of frequency of network oscillations similar to the slowing-down of the alpha rhythm in Alzheimer's disease. However, the networks are vulnerable to targeted attacks. A removal of a few excitatory or inhibitory hubs can impair sustained network oscillations.
\end{abstract}

\maketitle

\section{Introduction}
\label{intro}

Recent investigations have revealed that, on the functional level, brain has a complex network structure with small-world properties and scale-free architecture  \cite{sporns04,eguiluz05, honey07,bullmore09,sporns11}.
This kind of structure has been found in many other real physical, biological, and social systems where it plays an important role in dynamics and function
of the systems \cite{barabasi99,dg2002,newman2003,Boccaletti2006,dorogovtsev08,caldarelli07,Castellano09}.
%
It has been found that scale-free networks describing a number of complex systems, such as the World Wide Web, Internet, social networks or a cell, demonstrate error tolerance and attack vulnerability \cite{albert00,cohen00,cohen01}.
Networks with divergent second moment of degree distribution, such as scale-free networks with degree distribution $P(q)\propto q^{-\gamma}$ when $2 < \gamma \leq 3$, are more robust against random damage in comparison to networks with a finite second moment of degree distribution, such as scale-free networks with degree exponent $\gamma > 3$ or classical random graphs.
It is practically impossible to damage the giant connected component in an infinite uncorrelated network with divergent second moment of degree distribution. This kind of network is ultraresilient against random damage
or failures \cite{albert00,cohen00}. In contrast to random damage, the removal of highly connected hubs from a scale-free network effectively destroys the connectivity \cite{albert00,cohen01,callaway00}.
A close relationship of structural and functional neuronal networks and the importance of topological considerations for the function of the brain are discussed in reviews \cite{bullmore09,meunier10,sporns11}.
Theoretical investigations of neuronal networks revealed that scale-free structure may play an important role in synchronous neuronal activity \cite{batista07,grinstein05},
neuronal avalanches \cite{pellegrini07},  and neurological diseases \cite{morgan08}. An important feature of scale-free networks is a large number of hubs in comparison with classical random graphs with the same mean degree. Hubs may play an important role in connectivity,
promoting hyperexcitability after brain injury \cite{morgan08}, providing mechanism for orchestrating synchrony \cite{bonifazi09} or integrating multisensory information \cite{zamoralopez10}.
Recently, a "rich club" of densely interconnected hub regions was found in the human brain \cite{heuvel11,Sporns2012} similar to "rich club" found in other real complex systems \cite{zhou04,colizza06}.  In the brain, this rich club forms a central
backbone for global brain communication \cite{heuvel11,Sporns2012}.
Connectivity plays a critical role in mediating cognitive
function. Neurodegenerative diseases
target specific and functionally connected neuronal networks \cite{palop10syndep, palop10amyloid}.
Alzheimer's disease is characterized by the loss of neurons and synapses in the cerebral cortex and certain subcortical regions. This results in a slowing down of the alpha rhythm up to disappearance of this oscillation in the severe stages \cite{rodriguez11}. At the present time, understanding of a role of the scale-free structure and hubs in brain dynamics and mechanisms of the impairment of brain functions caused by damage of neuronal networks is elusive.


Besides the fact that the brain has a heterogeneous structure of complex networks, brain is noisy.
Usually, it is assumed that increase of noise level leads to increasing disorder.
However, under certain conditions,  noise can play a
constructive role, leading to stochastic resonance and
enhancement of weak signals
\cite{gammaitoni98,wm95} or generating coherent behavior \cite{bpt94,zhou05,springer06}.
Various sources of noise
were identified in the brain such as intracellular and synaptic noise and many others
\cite{Calvin67,white00,faisal08}.
A role of noise in brain activity was already discussed in a large
body of literature in neuroscience, see recent reviews \cite{white00,faisal08,ermentrout08}. In the brain,
noise stimulates spontaneous neuronal activity, fluctuation phenomena, and stochastic effects \cite{faisal08,lgns04}. It strongly influences both integrative properties of individual neurons \cite{white05,rapp92,destexhe99,ho00,destexhe03,destexhe10} and collective neuronal activity \cite{amit97,brunel00,brunelwang03,izhikevich08}.
Taking into account the fact that noise affects brain dynamics, it is important to understand the following: Does complex network architecture of neuronal networks play a role in the impact of noise on the brain activity?



In the present paper, we study the role of the scale-free architecture and noise
in activity of neuronal networks.
For this purpose, we use a cortical circuit model \cite{goltsev10} in which neurons form a strongly heterogeneous network with scale-free structure. In the considered networks, there are many excitatory and inhibitory hubs that are mutually and densely interconnected.
By use of numerical simulations and analytical considerations of the cortical model, we show that noise strongly enhances spontaneous activity of neuronal networks with scale-free structure.
In the case of scale-free networks with divergent second moment of degree distributions of synaptic connections, even a weak noise can stimulate spontaneous asynchronous activity of a finite fraction of neurons and sustained network oscillations.
Furthermore, we show error tolerance of scale-free neuronal networks to random damage in a wide range of the number of randomly removed excitatory or inhibitory neurons. Only damage of a substantial part of the network
(about $56\%$ of excitatory neurons or about $25 \% $  of inhibitory neurons for the networks studied in the paper)
impairs synchronization between neurons and suppresses sustained network oscillations. We demonstrate that excitatory and inhibitory hubs
play a crucial role in orchestrating neuronal activity.
Targeted attack on hubs effectively impacts collective behavior of neuronal networks. The removal of only a few hubs can produce an effect as strong as random removal of a finite fraction  of neurons. In particular, we find that removal of even a small number of hubs decreases the frequency of network oscillations and ultimately suppresses sustained network oscillations.

\section{Cortical circuit model}
\label{model}

In the present paper, we study a cortical circuit model of neuronal networks composed of stochastic excitatory and inhibitory neurons \cite{goltsev10}. The advantage of this model is that it includes biologically important constituent elements of real neuronal networks, it does not need time- consuming simulations, it takes explicitly  into account network heterogeneity, and it can be solved analytically. The analytical consideration enables us to understand mechanisms and origin of collective phenomena and the role of finite size effects in dynamics of neuronal networks. Our approach is based on assumption that there are universal phenomena in collective dynamics of neural networks that are independent on individual dynamics of neurons and depend on topological structure of the network and interaction between neurons. Investigation of these universal phenomena is the main aim our paper.

First, in contrast to models with integrate-and-fire neurons \cite{amit97,brunel00,brunelwang03} or another kind of individual dynamics \cite{izhikevich03,izhikevich06,izhikevich08},
we consider stochastic neurons like those of \cite{goltsev10,Cowan_1968,Benayoun_2010,Wallace_2011}. In this approach, a response of neurons on input is stochastic. A sufficiently large input
activates a neuron with a certain rate as in the Hopfield model \cite{hopfield82}. 

Second, we assume that excitatory and inhibitory neurons are tonic. They
fire spikes with a constant frequency $\nu$ for any input larger than a threshold.
In this respect, neurons
behave like those of \citet{McCulloch_1943}. The firing frequency $\nu$ is the same for both excitatory and inhibitory neurons.
Spikes mediate interactions between neurons. Total input from presynaptic neurons consists of spikes from excitatory neurons that activate a postsynaptic neuron and spikes from inhibitory neurons that suppress activity of the postsynaptic neuron. The total input $V_{j}(t)$ at neuron $j$ at time $t$ can be written as sum of contributions from excitatory and inhibitory neurons arriving during the time interval $[t-\tau,t]$ where $\tau$ is the integration time:
\beq
\label{eq:inputvsthreshcompar}
V_{j}(t)= \sum_{\ell=1}^{N} J_{\ell j} n_{\ell}(t)
\eeq
where $J_{\ell j}$ is the efficacy of the synaptic connection between pre- and postsynaptic neurons $\ell$ and  $j$, respectively.
In the case $\nu \tau > 1$, we assume that the number $n_{\ell}(t)$ of spikes generated by active neuron $\ell$ during time $\tau$ is given by $n_{\ell}(t)=\nu \tau $. If $\nu \tau < 1$, then  $n_{\ell}(t)=1 $ with the probability $\nu \tau$, otherwise, $n_{\ell}(t)=0 $ \cite{goltsev10}.
In the present paper, for simplicity, we will consider the case $\nu \tau > 1$. The case $\nu \tau < 1$ leads to qualitatively similar results \cite{goltsev10}. Inactive neurons give no contribution to $V_{j}(t)$. Furthermore, we assume that all efficacies of synaptic connections of excitatory neurons to postsynaptic excitatory and inhibitory neurons are uniform and positive, $J_{\ell j}=J_e$.
Efficacies of synaptic connections of inhibitory neurons to postsynaptic excitatory and inhibitory neurons are also uniform but negative and equal to $J_i$.

Third, in the considered model, we take into account noise.
The noise level is characterized by rates $f_e$ and $f_i$ that determine the probabilities $\tau f_e$ and $\tau f_i$ that an excitatory or inhibitory neuron, respectively, is activated by noise during time interval $\tau$. In our model, this stochastic process of activation of neurons represents action of synaptic noise (see Appendix \ref{sec:approxestim}).

\subsection{Static model for scale-free neuronal networks}
\label{sec:structure}

In order to study scale-free neuronal networks
we use the static model \cite{goh01, lee04, catanzaro05} that was developed for random uncorrelated scale-free complex networks. We generalize the model and build scale-free directed networks with two populations of neurons.
The model allows us to build a strongly heterogeneous network with both excitatory and inhibitory hubs. These hubs are mutually and densely interconnected.

\begin{figure}[b]
\bc
\includegraphics[height=6cm, angle=0]{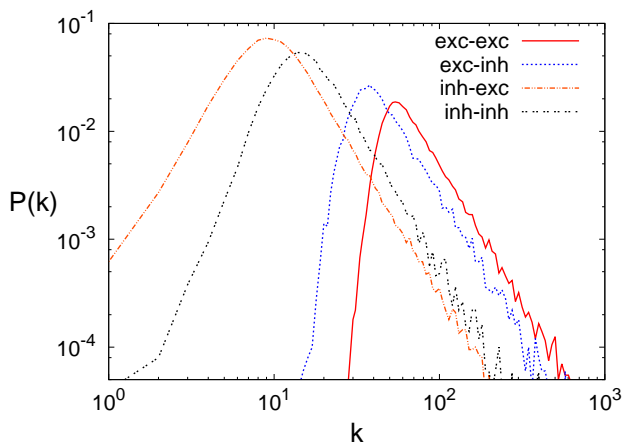}
\ec
\vspace*{-0.6cm}
\caption[]{\small \label{fig:degrdistsingleaverage}
Degree distributions of presynaptic connections
of excitatory and inhibitory neurons  in the static model Eq.~(\ref{eq:palbj}).
The degree distributions of excitatory-excitatory, excitatory-inhibitory, inhibitory-excitatory,
and inhibitory-inhibitory directed synaptic connections are obtained by averaging over 25 network
realizations for the following parameters: $N=10000$; $g_i=0.2$; $\gamma=2.5$; $K_{ee}=K_{ii}=150$; $K_{ei}=K_{ie}=100$.
}
\end{figure}

Let us consider a network composed by $N_e$ excitatory and $N_i$ inhibitory
neurons. They are numerated by indices $i=1,2,\dots,N_e$ and $j=1,2,\dots,N_i$, respectively. The fraction of excitatory neurons is $g_e =N_e/N$ and $g_i =1-g_e$ is the fraction of inhibitory neurons where $N=N_e +N_i$ is the total number of neurons. The weight
\beq
\label{eq:populweights}
w_a(j)=\frac{j^{-\lambda_a}}{\sum_{j =1}^{N_a}j^{-\lambda_a}}
\eeq
is assigned to each neuron $j=1, \ldots, N_a$  in population $a=e,i$. The weights are normalized, i.e., $\sum_{j =1}^{N_a} w_a(j)=1$.
We consider the case
$\lambda_e =\lambda_i \equiv \lambda$ with $\lambda \in [0,1)$.
These weights are used to construct a directed network.
The probability to have a directed connection from neuron $\ell$ in population $a$ to a neuron $j$ in population $b$ is
\beq
\label{eq:palbj}
p_{a,\ell;b,j}\equiv N g_a K_{ab} g_b  w_a(\ell)w_b(j) \; .
\eeq
According to this probability and the weight Eq.~(\ref{eq:populweights}), a neuron in population $b$ with a small index $j$ has a larger probability to have a connection with neurons in populations $a$ in comparison with a neuron with a larger index $j$. Therefore, neurons with a small index $j$ have a large probability to become hubs.
Neurons in population $b$ have in average $C_{ab}$  presynaptic neurons in population $a$, where
\beq
\label{eq:mean-d}
C_{ab}\equiv\frac{1}{N_b}\sum_{j =1}^{N_b} \sum_{\ell =1}^{N_a} p_{a,\ell;b,j}= g_a  K_{ab}.
\eeq
An excitatory neuron has in average $C_{ee}$ of presynaptic excitatory and $C_{ie}$ presynaptic inhibitory neurons. An inhibitory neuron has in average $C_{ei}$ of presynaptic excitatory and $C_{ii}$ presynaptic inhibitory neurons. This determines the role of the matrix $K_{ab}$.

The in-degree distribution
in a scale-free network built by use of the probability Eq.~(\ref{eq:palbj}) is shown in Fig.~\ref{fig:degrdistsingleaverage}.
One can see that the degree distribution is a power law at a sufficiently large degree $k$ ,
\beq
P(k)\sim k^{-\gamma},
\eeq
with the degree exponent
\beq
\gamma=1+\frac{1}{\lambda} \,\,.
\eeq
Using this model, one builds a network with $\gamma > 2$. Note that, in the static model, in- and out-degrees are correlated
because the weight Eq.~(\ref{eq:populweights}) determines the probability Eq.~(\ref{eq:palbj}) of both pre- and postsynaptic connections. When building a network, we excluded multiple and self-links. In the case $\gamma < 3$, the exclusion of multiple links causes degree-degree correlations, namely,  disassortative mixing \cite{goh01, lee04, catanzaro05}.

\subsection{Erd\"{o}s-R\'{e}nyi networks}
\label{ER graph}

If, in Eq.~(\ref{eq:palbj}), we suppose that the weights
$w_e(j)=1/N_e$ and $w_i(j)=1/N_i$, then we obtain the probability $p_{a,\ell;b,j}= K_{ab}/N$ that does not depend on the indices $\ell$ and $j$. Constructing a network with this probability, we obtain a sparsely connected network with the same mean numbers of connections between excitatory and inhibitory neurons as in the corresponding scale-free network. The sparsely connected network has the Poisson degree distribution as the classical random or Erd\"{o}s-R\'{e}nyi networks. Comparing dynamics of
the static model and the corresponding randomly connected network, one can study a role of structure in network dynamics.

\subsection{Rules of stochastic dynamics. Algorithm}
\label{sec:rules}

In the cortical model,
sporadic inputs to neurons generate neuronal activity. Dynamics of neuronal populations is determined by the following rules:
\begin{itemize}
\item[(i)]
During the integration time $\tau$, inactive excitatory and inhibitory neurons are activated by noise with probabilities $\tau f_e$
and $\tau f_i$, respectively.
\item[(ii)]
An excitatory or inhibitory neuron is activated with probability $\tau \mu_{e}$ or $\tau \mu_{i}$, respectively, by an input $V(t)$,
if $V(t)$ is larger then a threshold $V_{th}$.
\item[(iii)]
Excitatory and inhibitory neurons stop firing with probabilities $\tau \mu_{e}$ and $\tau \mu_{i}$, respectively, if input $V(t)$ becomes smaller than $V_{th}$.
\end{itemize}
For simplicity, we assume that the threshold $V_{th}$ is the same for all neurons.

In the case $\nu \tau \geq 1$, the input Eq. (\ref{eq:inputvsthreshcompar}) takes a form, $V(t)=\nu \tau J_e n +\nu \tau J_i m$, where $n$ and $m$ are the number of active presynaptic excitatory and inhibitory neurons at time $t$, respectively.
It is convenient to introduce a dimensionless activation threshold $\Omega$,
$\Omega \equiv V_{th}/(J_{e} \nu \tau)$. The activation condition $V(t) \geq V_{th}$ takes a form, 
\beq
\label{activation}
n -|J_i/J_e| m \geq \Omega.
\eeq
Below we use the unit $J_e\equiv 1$.
$\Omega$ is of the order of
15-30 in living neuronal networks \cite{Eckmann07,breskin06,Soriano08} and about $30-400$ in the brain.

We assume that $1/\mu_{e}$ and $1/\mu_{i}$ are of the order of the first-spike latencies of excitatory and inhibitory neurons, respectively.
First-spike latency is defined as the time from the onset of stimulus
to the time of occurrence of the first-response spike \cite{Heil04}. It is about 6-7 ms in fast-spike interneurons in neocortex \cite{Swadlow03}.  The first spike latency is ranged from 25 to 49 ms in CA3 hippocampal pyramidal (excitatory) neurons \cite{fmi04} and 20 -128 ms in inhibitory cerebellar stellate cells \cite{mfmt05}. In our model, the ratio
\beq
\label{eq:alpha}
\alpha \equiv \frac{\mu_{i}}{\mu_{e}}
\eeq
is an important parameter that determines collective dynamics of neurons. According to the  experimental data, $\alpha$ may be both larger and smaller than 1.

In simulations, we apply the following algorithm. We divide time $t$ into intervals of width $\Delta t=\tau$.
At each time step, an inactive excitatory or inhibitory neuron may be activated by noise with probability $\Delta t f_e$ or $\Delta t f_i$, respectively.
Furthermore, at each time step, for each neuron, we calculate the input  Eq.~(\ref{eq:inputvsthreshcompar}), taking into account that, in the case $\tau \nu \geq 1$, each active presynaptic neuron contributes with $\tau \nu$ spikes.
In the case $\tau \nu < 1$, an active presynaptic neuron contributes with a spike with probability $\tau \nu$.
Then, with the probability $\Delta t \mu_a$, $a=e,i$, we update the states of all neurons using the stochastic rules formulated above. In our simulations, we only consider the case $\tau \nu \geq 1$. Throughout this paper we use $1/\mu_{e}\equiv 1$ as time unit.

\subsection{Rate equations for neuronal activity}
\label{sec:analytdyn}

\begin{figure}[h]
\bc
\includegraphics[height=5.32cm, angle=0]{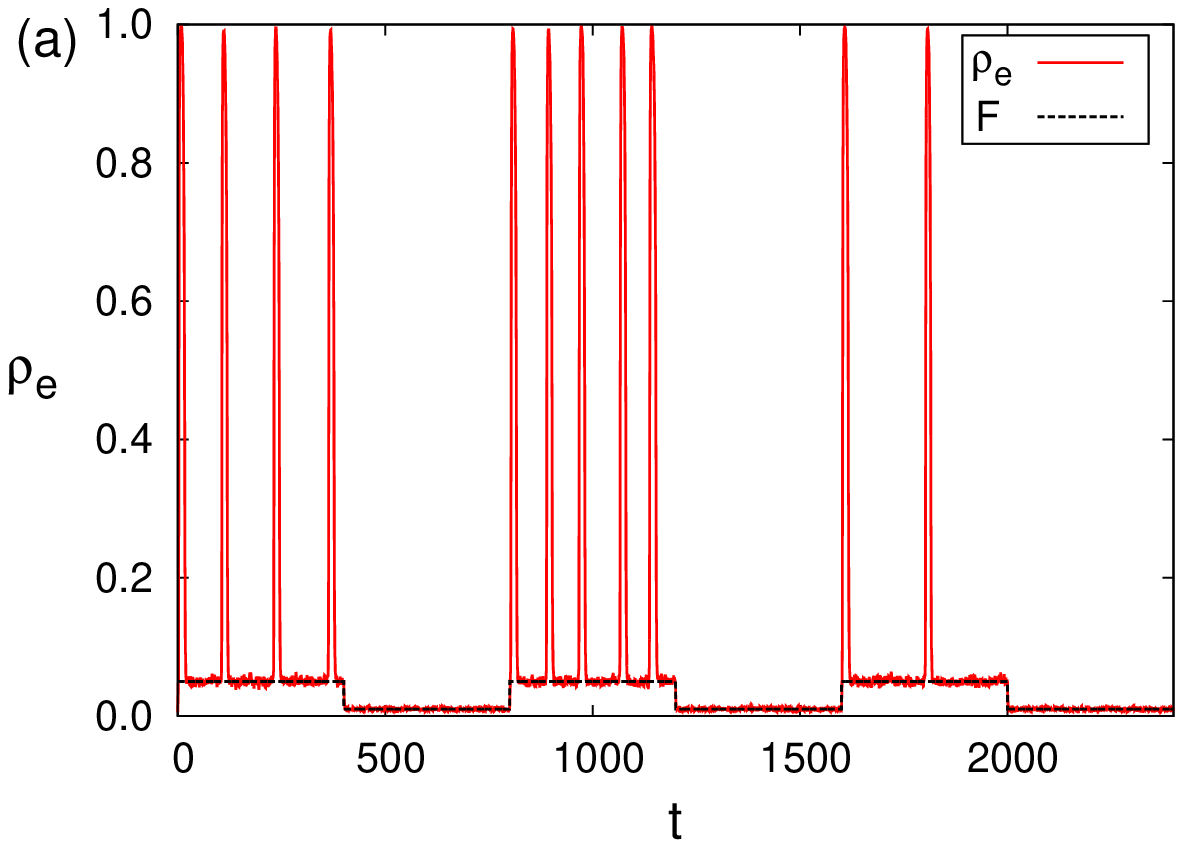} \\
\hspace*{-0.3cm}
\includegraphics[height=5.2cm, angle=0]{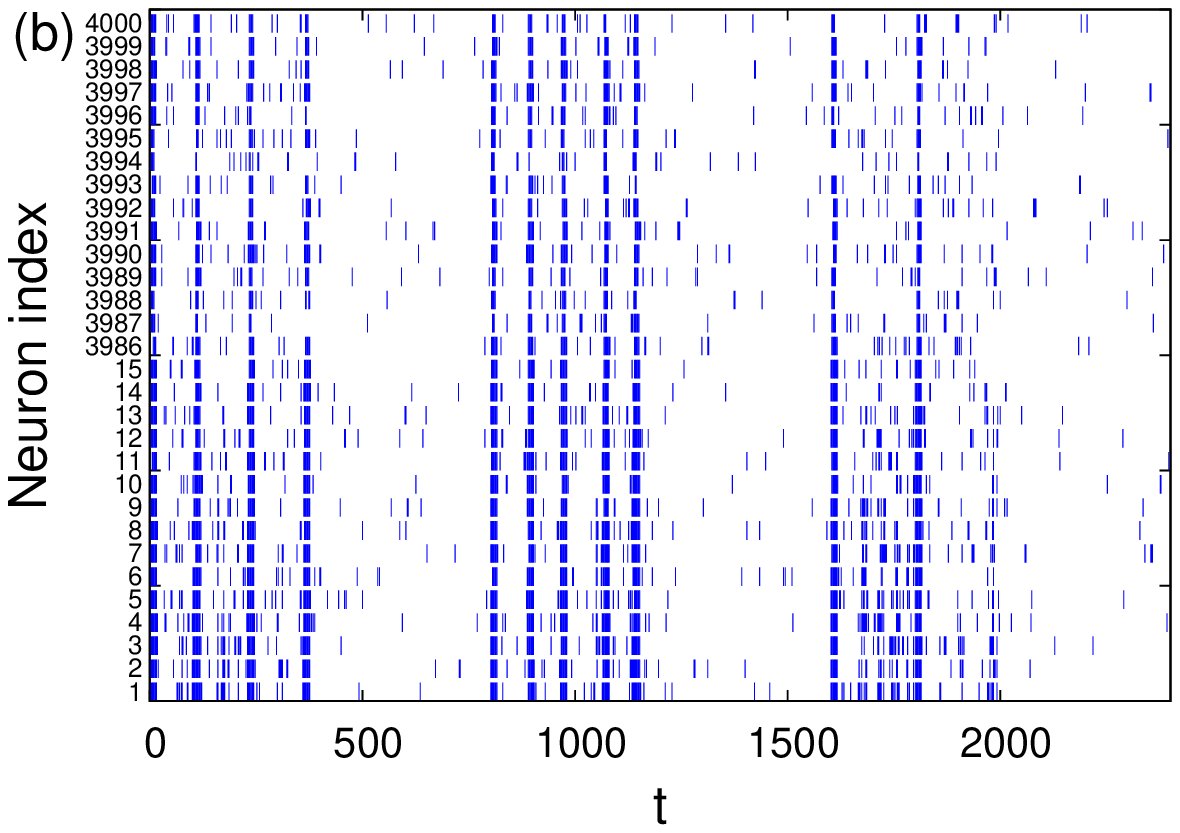}
\ec
\vspace*{-0.5cm}
\caption[]{\small\label{fig:rasterplotexamp}
Asynchronous behavior and sustained network oscillations in the cortical model with scale-free structure. (a) Periodic change of noise level $F$ leads to periodic change of activity $\rho_e$ of excitatory neurons: a low neuronal activity at a low noise level and sustained network oscillations at a high noise level. Inhibitory neurons (not shown) demonstrate similar behavior.
(b) Raster plot displays temporal behavior of individual excitatory neurons with index  $1,\dots, 4000$.
At a low noise level, the neuronal activities are weak and asynchronous.
At a high noise level, the neurons are synchronized and sustained network oscillations appear.
Note that neurons with a small index are hubs and have stronger activity in comparison to neurons having a large index and, therefore, a smaller number of synaptic connections.
Other parameters:
$N= 5000$, $g_i=0.2$, $\gamma=2.5$,
$K_{ab}=400$,
$J_{i}=-3.2$,
$\Omega=60$,
$\alpha=0.1$,
$\Delta t=0.1$.
}
\end{figure}

Let us study dynamics of neuronal networks with scale-free structure of the static model.
We simulated the model using the algorithm described above in Sec.~\ref{sec:rules}.  The raster plot in Fig.~\ref{fig:rasterplotexamp} represents activity of neurons in the case when the noise level is periodically changed. One can see that if the noise level is low, neurons show asynchronous behavior while at a high noise level their activities are correlated and form sustained network oscillations. In order to understand origin of collective behavior and mechanism of synchronization in neuronal networks, we develop a theoretical approach to dynamics of the cortical model.

Let us introduce the probability $\rho_{a}(\ell,t)$ that at time $t$ neuron $\ell$ in population $a$ is active. We define the activity $\rho_a(t)$ of population $a$ as follows:
\beq
\label{eq:rhoa}
\rho_a(t) \equiv \frac{1}{N_a}\sum_{\ell=1}^{N_a} \rho_{a}(\ell,t)
\eeq
The activity $\rho_a(t)$ represents neuronal activity in EEG measurements.
We also introduce a weighted activity $\tilde{\rho}_a(t)$ of population $a$,
\beq
\label{eq:rhotildeb}
\tilde{\rho}_a(t) \equiv \sum_{\ell=1}^{N_a} w_a(\ell) \rho_{a}(\ell,t).
\eeq
Based on the rules of the stochastic dynamics in Sec.~\ref{sec:rules} and the method developed in Ref. \cite{goltsev10}, one derives rate equations for the weighted activities of excitatory and inhibitory populations (see Appendix \ref{derivation}):
\begin{align}
\label{eq:eqofmotactivit}
& {\dot{\tilde{\rho}}}_e=f_e-\nu_e \tilde{\rho}_e+\mu_{e}\tilde{\Psi}_e(\tilde{\rho}_e, \tilde{\rho}_i), \nonumber \\
& {\dot{\tilde{\rho}}}_i=f_i-\nu_i \tilde{\rho}_i+\mu_{i}\tilde{\Psi}_i(\tilde{\rho}_e, \tilde{\rho}_i) ,
\end{align}
where ${\dot{\tilde{\rho}}}\equiv d\tilde{\rho}/dt$, $\nu_a \equiv f_a+\mu_{a}$, and
\beq
\label{eq:psitildeb}
\tilde{\Psi}_a(\tilde{\rho}_e, \tilde{\rho}_i)=\sum_{j=1}^{N_a}w_a(j) \, \Psi_{a,j}(\tilde{\rho}_e, \tilde{\rho}_i) \; .
\eeq
The function $\Psi_{a,j}(\tilde{\rho}_e, \tilde{\rho}_i)$ is the probability that, at given weighted activities $\tilde{\rho}_e$ and $\tilde{\rho}_i$, an input at neuron $j$ in population $a$ is at least the threshold $\Omega$,
\begin{align}
\label{eq:psibj}
\Psi_{a,j}(\tilde{\rho}_e, \tilde{\rho}_i)&=\sum_{n=\Omega}^{\infty}\sum_{m=0}^{\infty} P_{n}(\tilde{\rho}_{e}C_{ea}(j))P_{m}(\tilde{\rho}_{i}C_{ia}(j))\times \nonumber \\
& \Theta(n-|J_{i}|m-\Omega).
\end{align}
The product $P_{n}(\tilde{\rho}_{e}C_{ea}(j))P_{m}(\tilde{\rho}_{i}C_{ia}(j))$ is the probability that, at given weighted activities $\tilde{\rho}_e$ and $\tilde{\rho}_i$, neuron $j$ has $n$ active presynaptic excitatory and $m$ active presynaptic inhibitory neurons.  The Heaviside step function $\Theta(n-|J_{i}|m-\Omega)$ selects events when the input at a neuron $j$ is at least the threshold $\Omega$, Eq. (\ref{activation}). The function $P_n(x)=x^{n}e^{x}/n!$ is the Poissonian distribution function. $C_{eb}(j)$ and $C_{ib}(j)$ are the averaged number of excitatory and inhibitory presynaptic connections of neuron $j$ in population $b$,
\beq
\label{eq:Cabj}
C_{ab}(j) \equiv \sum_{\ell=1}^{N_a}p_{a,\ell;b,j} =N g_a K_{ab} g_b w_b(j).
\eeq
Solving Eqs.~(\ref{eq:eqofmotactivit}) with respect to $\tilde{\rho}_e(t)$ and $\tilde{\rho}_i(t)$, then one finds the activities $\rho_e(t)$ and $\rho_i(t)$, using equations
\begin{align}
\label{eq:EOMactivit}
& {\dot {\rho}}_e=f_e-\nu_e \rho_e+\mu_{e} \Phi_e(\tilde{\rho}_e, \tilde{\rho}_i), \nonumber \\
& {\dot {\rho}}_i=f_i-\nu_i \rho_i+\mu_{i} \Phi_i(\tilde{\rho}_e, \tilde{\rho}_i) ,
\end{align}
where
\beq
\label{eq:phi}
\Phi_a(\tilde{\rho}_e, \tilde{\rho}_i)=\frac{1}{N_a}\sum_{j=1}^{N_a} \Psi_{a,j}(\tilde{\rho}_e, \tilde{\rho}_i) \; ,
\eeq
(see Appendix \ref{derivation}). Rate equations (\ref{eq:eqofmotactivit}) are asymptotically exact in the limit of large mean degree and large size $N \gg 1$. Moreover, it is assumed that the integration time $\tau$ is much smaller than the relaxation time or the period of sustained network oscillations.

It is convenient to introduce a dimensionless parameter,
\beq
F_a=\frac{f_a}{f_a+\mu_{a}} \;
\label{F}
\eeq
for $a=e,i$. $F_a$ characterizes the activation rate $f_a$ of neurons stimulated by noise with respect to the rate $\mu_{a}$. If $F \ll 1$, it corresponds to a small noise level, i.e.,  the probability of activation by noise is small (see Appendix \ref{sec:approxestim}).
For simplicity, we consider the case $F_e=F_i\equiv F$.

\begin{figure}[h]
\bc
\includegraphics[height=6cm, angle=0]{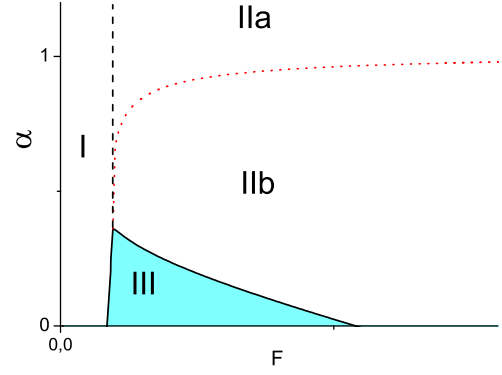}
\ec
\vspace*{-0.5cm}
\caption[]{\small\label{pd}
Schematic representation of the $\alpha - F$ plane of the cortical model. (I) A region with low activity of neurons, asynchronous behavior, and exponential relaxation to a steady state; (II) A region with
high activity of neurons.
The dotted line divides region II into two subregions: region IIa  with exponential relaxation and region IIb with damped network oscillations. (III) A region with sustained network oscillations. The vertical dashed line is the critical line of a first-order phase transition from a state with low neuronal activity to a state with high activity when $\alpha > \alpha_t$. Above the critical noise level only the state with a high neuronal activity is stable.
}
\end{figure}

\begin{figure}[h]
\bc
\includegraphics[height=6cm, angle=0]{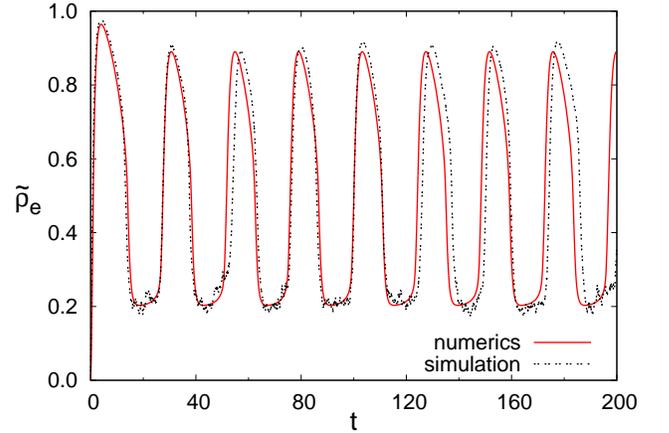}
\ec
\vspace*{-0.5cm}
\caption[]{\small\label{fig:numsimdyncompare}
Sustained network oscillations of activity of excitatory population in the cortical model from  a numerical solution of Eq.~(\ref{eq:eqofmotactivit}) (solid line) and our simulations (dotted line) for parameters:
$N=10000$, $g_i=0.2$, $\gamma=2.5$,
$K_{ab}=75$,
$F=0.2$,
$J_{i }=-3.5$,
$\Omega=10$,
$\alpha=0.1$,
$\Delta t=0.1$.
}
\end{figure}

Equations ~(\ref{eq:eqofmotactivit}) are non-linear and their explicit solution is unknown. We solved them numerically.
Analyzing a linear response of the activities on small perturbations of the noise level, we found that the phase diagram of the cortical model with scale-free structure is qualitatively similar to the phase diagram found in Ref.~\cite{goltsev10} for sparsely connected networks with Poisson degree distribution. In Fig.~\ref{pd}, there are three regions:  region I with low activity and asynchronous behavior of neurons and exponential relaxation to a steady state; region II with high activity of neurons and either exponential relaxation or relaxation in the form of damped oscillations to a steady state;  region III with sustained network oscillations.

In Fig.~\ref{fig:numsimdyncompare} we compare between a numerical solution of Eqs.~(\ref{eq:eqofmotactivit}) and simulations of the cortical model of size $N=10000$ with degree exponent $\gamma=2.5$.
One can see that shape and
frequency of sustained network oscillations calculated from Eqs.~(\ref{eq:eqofmotactivit}) are in good agreement with simulations. This result shows that Eqs.~(\ref{eq:eqofmotactivit}) actually describe correctly the dynamics of the cortical model.

\section{Noise-induced spontaneous activity and sustained oscillations}

In Sec.~\ref{intro} we have noted that noise plays an important role in activity of neuronal networks.
In this section we will show that the level of noise necessary to sustain activity of a finite fraction of neurons depends on the network structure. In scale-free networks with fat tailed degree distribution, it can be very small.

\subsection{Spontaneous activity in scale-free networks}
\label{subsec:depactnoise}

\begin{figure}[h]
\bc
\includegraphics[height=6cm, angle=0]{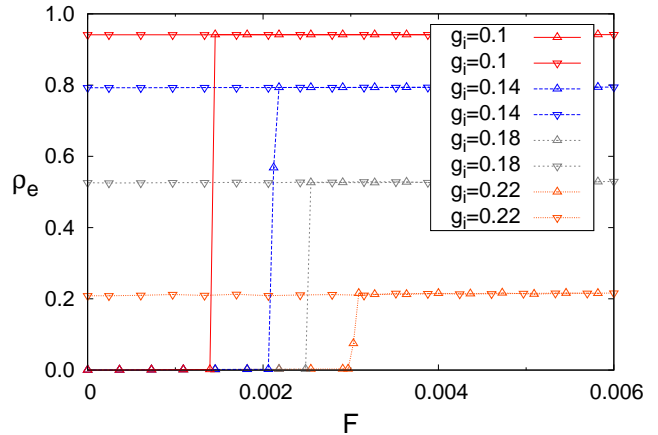}
\ec
\vspace*{-0.4cm}
\caption[]{\small
\label{fig:activationplot_N50000_gi0k1_0k02_ise2k4_nrnetw6_degreeboth}
Excitatory activity $\rho_e$ versus the noise level $F$ at different fractions $g_i$ of inhibitory neurons in cortical model of neuronal networks with scale-free structure. With increasing $F$ 
(triangles up, $\bigtriangleup$) a discontinuous activation transition occurs at a critical noise level $F_c$. With decreasing $F$
(triangles down, $\bigtriangledown$) the network remains in the metastable active state.
$N= 50000$, $\gamma=2.5$,
$K_{ab}=100$,
$J_{i}=-2.4$,
$\Omega=18$,
$\alpha=1$,
$\Delta t=0.1$.
}
\end{figure}

Even in the absence of any stimulus, neuronal networks demonstrate spontaneous activity that may be generated by noise. Let us study how spontaneous activity depends on structure of neuronal networks. For this purpose, we consider the static model of scale-free neuronal networks from Sec.~\ref{sec:structure}.
First we consider the case when $\alpha$ is larger than a certain value $\alpha_t$ above which there are no sustained network oscillations ($\alpha_t$ is the $\alpha$ coordinate of the top point of the region III in Fig.~\ref{pd}).
In this case, with increasing the noise level $F$ the activity of the neuronal network undergoes a discontinuous transition with a jump in excitatory and inhibitory activity at a critical noise level $F_c$, see  Fig.~\ref{fig:activationplot_N50000_gi0k1_0k02_ise2k4_nrnetw6_degreeboth}. In the schematic phase diagram in Fig.~\ref{pd}, the critical noise level $F_c$ determines the line of the first-order phase transition. At a low noise level $F < F_c$, the activities are very small because neurons are activated asynchronously and rarely by noise. When $F$ approaches $F_c$, one observes avalanches stimulated by interaction between neurons \cite{goltsev10}. Avalanches in real neuronal networks have been revealed by Beggs and Plenz \cite{Beggs_2003}, see also recent reviews \cite{Plenz_2007,Chialvo_2006,Chialvo_2010}. A detailed discussion of avalanches is out of the scope of the present paper.

Another interesting phenomenon is hysteresis.
%
Starting from a weak noise, at first we increased the noise level $F$ from zero to a value above the critical point $F_c$ and afterwards decreased it again to a value below $F_c$. In Fig.~\ref{fig:activationplot_N50000_gi0k1_0k02_ise2k4_nrnetw6_degreeboth} one can see that the network activity remains as high as it was above $F_c$.
In our simulations, we found that the neuronal network stays in this metastable active state for a long time because the activity is supported by spiking neurons. A similar phenomenon was found in a model of mammalian thalamocortical systems in Ref. \cite{izhikevich08} where spiking dynamics of each neuron was simulated by use of Izhikevich  model \cite{izhikevich03}. The authors \cite{izhikevich08} observed that the neuronal activity remained high even after turn-off the spontaneous synaptic release that was a driving force of initial activity.

\subsection{Critical level of noise}
\label{subsec:FcvsN}

\begin{figure}[h]
\bc
\includegraphics[height=6cm, angle=0]{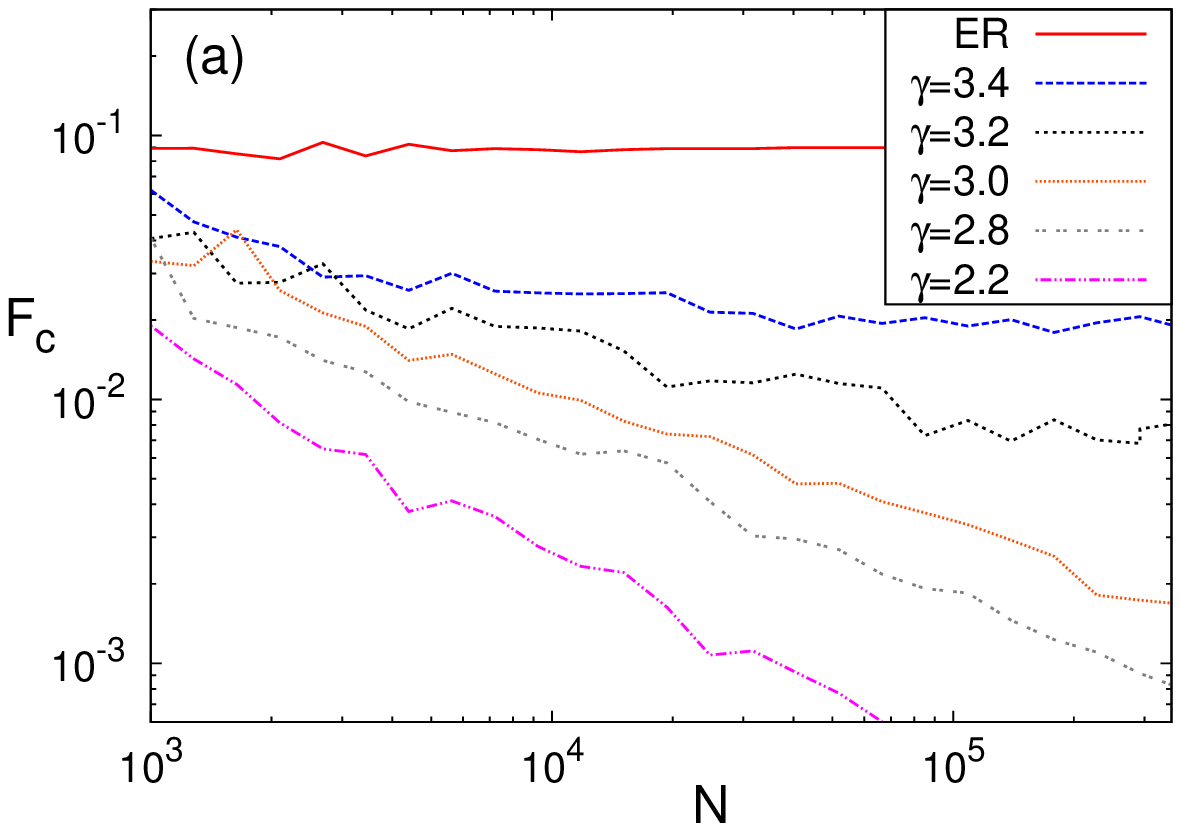}
\includegraphics[height=6cm, angle=0]{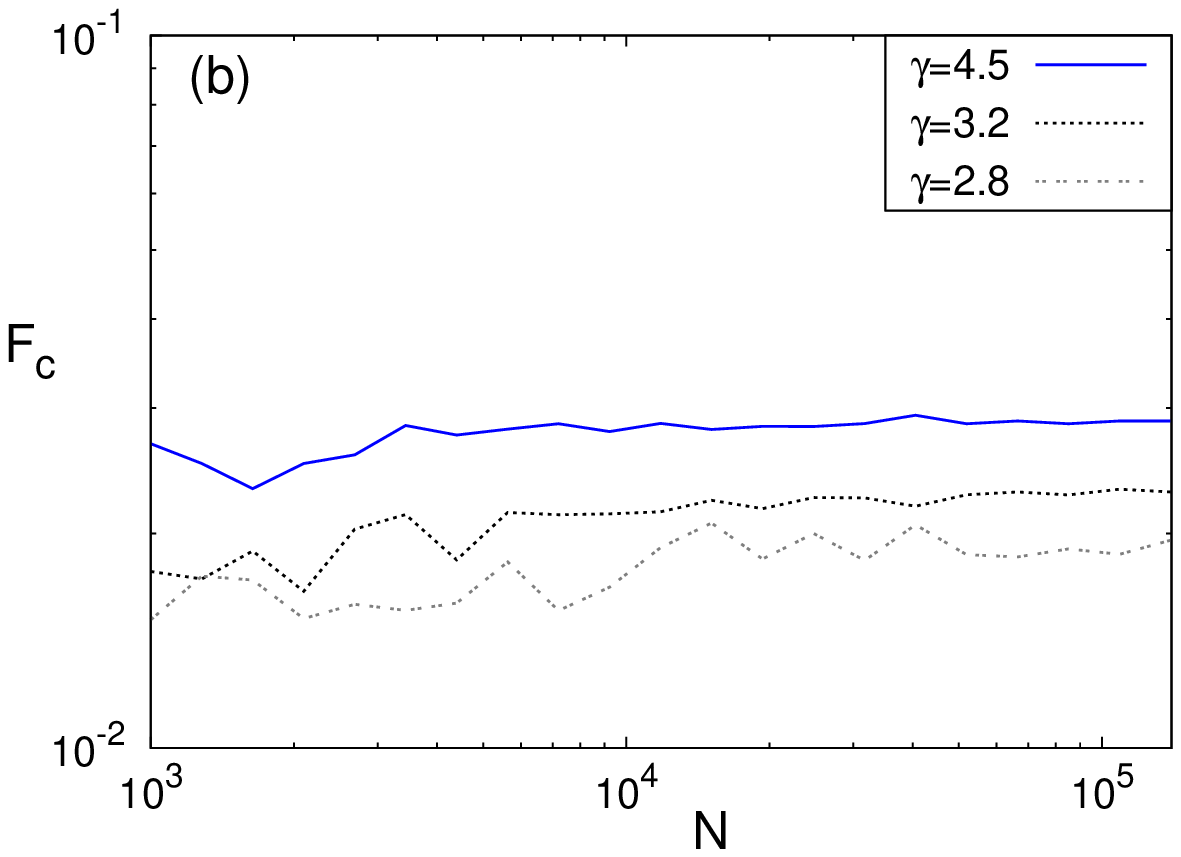}
\ec
\caption[]{\small\label{fig:Fc_vs_N_gammavariab}
Critical noise level $F_c$ of the first-order phase transition versus network size $N$.
(a) Scale-free neuronal networks (static model).
(b) The same networks as in $(a)$ but with shuffled in- and out-degrees.
Each data point was obtained by averaging over 5 network realizations. Other parameters:
$g_i=0.2$,
$K_{ab}=100$,
$J_{i}=-3.0$,
$\Omega=12$,
$\alpha=1.0$.
}
\end{figure}

Our simulations revealed that the critical point $F_c$  of the discontinuous transition depends on both structure and size of the network. Our results are represented in Fig.~\ref{fig:Fc_vs_N_gammavariab}. In networks
with the degree exponent $2 < \gamma \leq 3$, such as scale-free networks revealed by fMRI in humans \cite{eguiluz05}, the critical noise level $F_c(N) $ decreases with increasing the size $N$ following a power law,
\beq
F_c(N) \propto N^{-\sigma}.
\eeq
The exponent $\sigma$ depends on the degree exponent $\gamma$: $\sigma(\gamma=2.8)\approx 0.7$ and $\sigma(\gamma=2.2)\approx 0.82$
(Fig.~\ref{fig:Fc_vs_N_gammavariab}a).
In the case of the Erd\"{o}s-R\'{e}nyi networks,
the critical noise $F_c$ of the discontinuous activation transition is finite and does not depend on size at $N \gg 1$, see Fig.~\ref{fig:Fc_vs_N_gammavariab}a. The Erd\"{o}s-R\'{e}nyi networks were constructed using the methods described in Section~\ref{ER graph}. Neurons in the networks have the same mean degrees as the corresponding scale-free networks.

In Appendix \ref{sec:analytcalcs}, solving Eqs.~(\ref{eq:eqofmotactivit}) for a steady state, ${\dot {\rho}}_e={\dot {\rho}}_i= 0$, we show that in the cortical model with the scale-free
degree distribution at $\gamma < 3$, in the limit $N \rightarrow \infty$,
the critical value $F_c$ of noise in the activation process tends to zero,
It means that the activation phase transition described in Sec.~\ref{subsec:depactnoise} takes place at $F_c=0$, i.e., at an arbitrary small noise level $F$ a finite fraction of the neuronal network is activated. However, $F_c \neq 0$ if $\gamma>3$ or $N$ is finite.

We would like to note that the absence of the threshold $F_c$
was obtained within the static model. In this model, in- and out degrees are correlated.
If a neuron has a large in-degree, then with high probability it also has a large out-degree. If these correlations are removed, for example, shuffling in- and out-degrees like in papers \cite{caldarelli07,morgan08}, then, according to our simulations, a finite threshold is restored,
though it is much smaller compared to the Erd\"{o}s-R\'{e}nyi networks, see Figs.~\ref{fig:Fc_vs_N_gammavariab}a and \ref{fig:Fc_vs_N_gammavariab}b.

\subsection{Noise stimulated network oscillations}

\begin{figure}[h]
\bc
\includegraphics[height=6.0cm, angle=0]{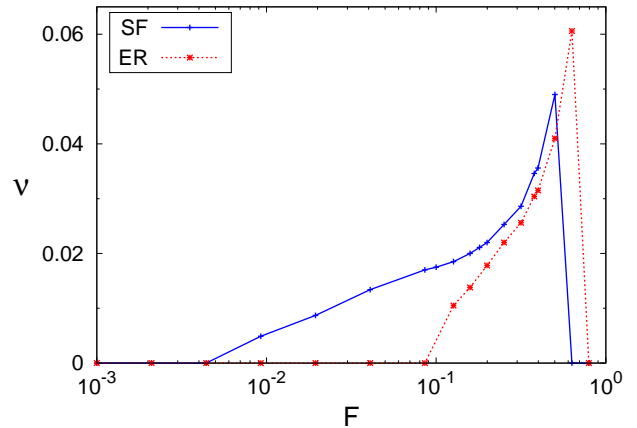}
\ec
\vspace*{-0.5cm}
\caption[]{\small\label{fig:dyn_and_freq}
Frequency $\nu$ of sustained network oscillations versus the noise level $F$ in a scale-free (SF) network (static model) and Erd\"{o}s-R\'{e}nyi (ER) network.
Parameters in simulations:
$N= 50000$, $g_i=0.2$,
$\gamma=2.5$ (static model),
$K_{ab}=2000$,
$J_{i}=-3.5$,
$\Omega=100$,
$\alpha=0.1$,
$\Delta t=0.1$.
}
\end{figure}

If the parameter $\alpha$, Eq.~(\ref{eq:alpha}), is smaller than a critical value $\alpha_t$ defined in Sec.~\ref{subsec:depactnoise}, then with increasing the noise level $F$, the cortical model undergoes a phase transition to a state with sustained network oscillations.
In Fig.~\ref{fig:dyn_and_freq}, we compare frequencies of sustained network oscillations induced  by noise in Erd\"{o}s-R\'{e}nyi and scale free neuronal networks having the same mean in- and out- degrees and constructed according to Sec. \ref{sec:structure} and \ref{ER graph}.
One can see that the frequency of the oscillations depends on both the noise level $F$ and the network structure. Furthermore, network oscillations in the scale-free networks appear at a much weaker noise than in the randomly connected network.

\section{Random damage and targeted attacks on neuronal networks}

Alzheimer's disease is characterized by the loss of neurons and synapses in the cerebral cortex and certain subcortical regions and it is accompanied by the slowing-down of the alpha rhythm \cite{rodriguez11}. In this section, motivated by these observations,  we study impact of random damage and targeted attacks on neuronal networks.

\subsection{Random damage of neuronal networks}

\begin{figure}[h]
\bc
\includegraphics[height=6cm, angle=0]{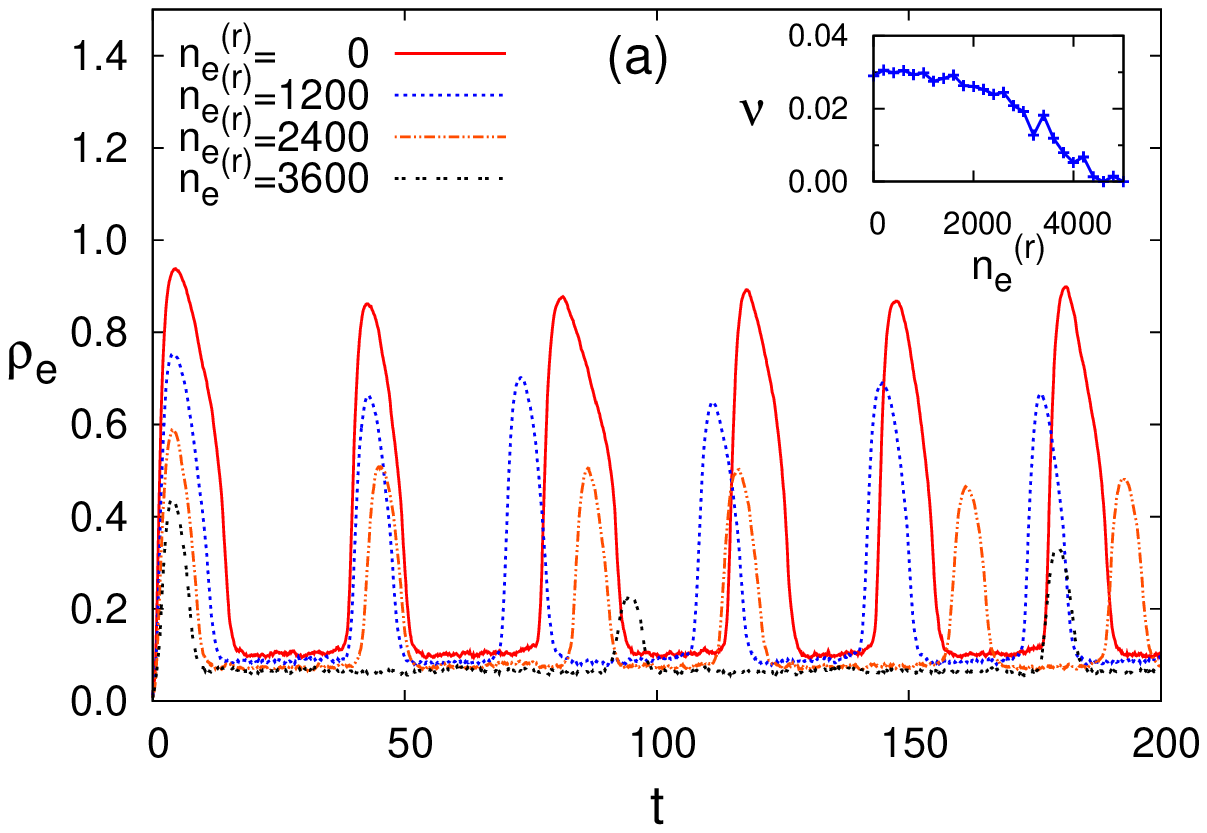}
\includegraphics[height=6cm, angle=0]{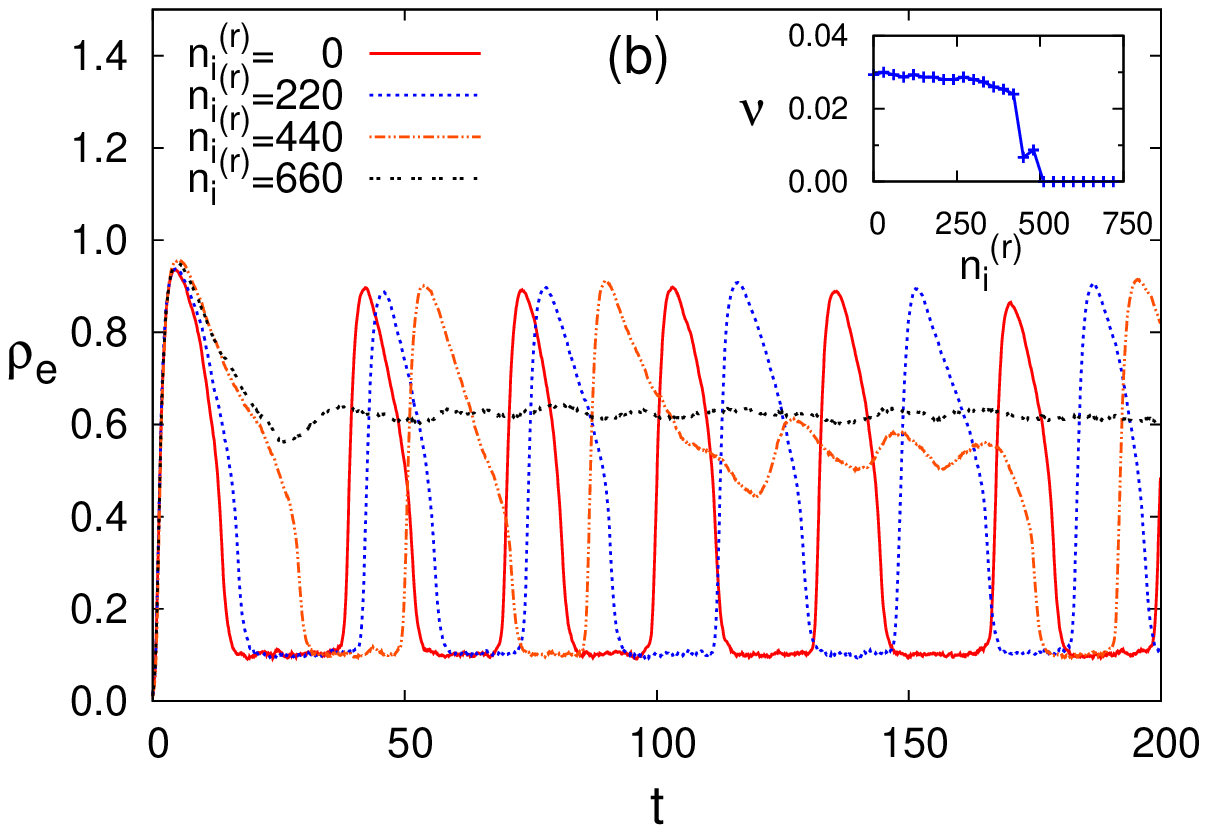}
\ec
\vspace*{-0.4cm}
\caption[]{\small\label{fig:randomremovalunshuffled}
Sustained network oscillations of excitatory activity in randomly damaged scale-free networks.
(a) Random removal of $n_e^{(r)}$ excitatory neurons, $n_e^{(r)}=0, 1200, 2400$, and 3600. The corresponding fractions of randomly removed excitatory neurons are  $Q_e=0, 0.15, 0.3$, and 0.45.
(b) Removal of $n_i^{(r)}$ inhibitory neurons, $n_i^{(r)}=0, 220, 440$, and 660. The corresponding  fractions of randomly removed inhibitory neurons are $Q_i=0, 0.11, 0.22$, and 0.33.
Insets: Frequency $\nu$ of sustained network oscillations versus the number of removed excitatory or inhibitory neurons. Other parameters:
$N= 10000$, $g_i=0.2$, $\gamma=2.5$, $K_{ab}=75$, $F=0.1$,
$J_{i}=-3.5$,
$\Omega=10$,
$\alpha=0.1$,
time step
$\Delta t=0.1$.
}
\end{figure}
We start by investigating the impact of random removal of excitatory or inhibitory neurons on synchronization and frequency of sustained network oscillations in neuronal networks similar to scale-free networks revealed by fMRI in humans \cite{eguiluz05}.
In simulations, we studied random damage of scale-free networks that consist of 10000 neurons (80\% excitatory and 20\% inhibitory neurons) and have a fat-tailed degree distribution with $\gamma=2.5$ (the static model from Sec.~\ref{sec:structure}).
In Fig.~\ref{fig:randomremovalunshuffled}, one can see
that the sustained network oscillations are tolerant to errors in sufficiently broad ranges of the fractions $Q_e$ and $Q_i$ of randomly removed excitatory or inhibitory neurons. With increasing
$Q_e$ and $Q_i$, the amplitude and frequency of sustained network oscillations decrease. Removal of about $56\%$ of excitatory neurons or about $25 \% $  of inhibitory neurons suppresses completely the oscillations. Thus, the network is more tolerant to random removal of excitatory neurons than to random removal of inhibitory neurons. The main mechanism of this effect is
decrease of the network’s connectivity that, in turn, attenuates synchronization between neurons. A similar decrease of power and frequency of alpha rhythm up to disappearance of alpha rhythms in the severe stages occurs in  Alzheimer's disease \cite{rodriguez11}.

Above we found that scale-free neuronal networks display a surprisingly high degree of tolerance
against random failures.
A similar effect occurs if, instead of neurons, a certain fraction of synapses is randomly removed. This assumption is based on the fact that, in complex networks, bond and site percolation have similar properties  \cite{callaway00}. Robustness of scale-free neuronal networks to damage may be even enhanced if the networks have community structure \cite{wu09}.

\subsection{Vulnerability of scale-free neuronal networks to targeted attacks}

\begin{figure}[h]
\bc
\includegraphics[height=6cm, angle=0]{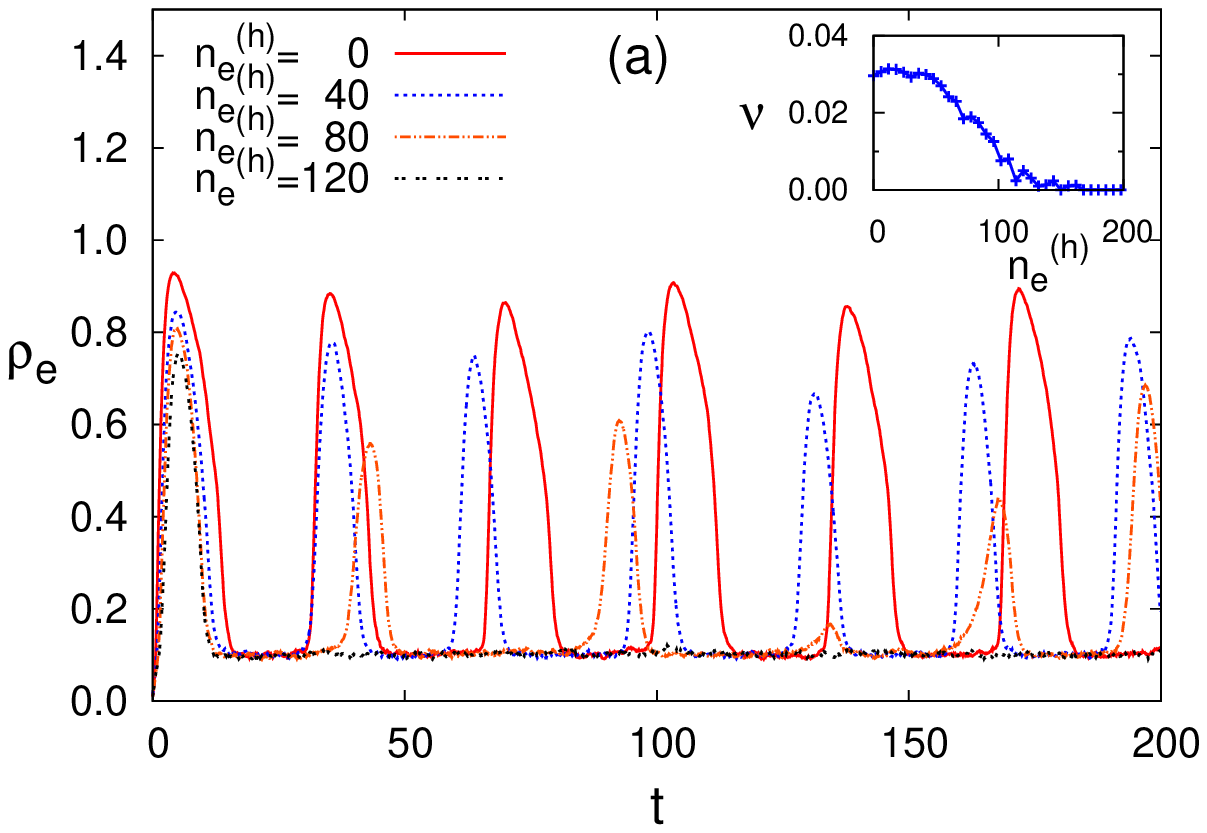}
\includegraphics[height=6cm, angle=0]{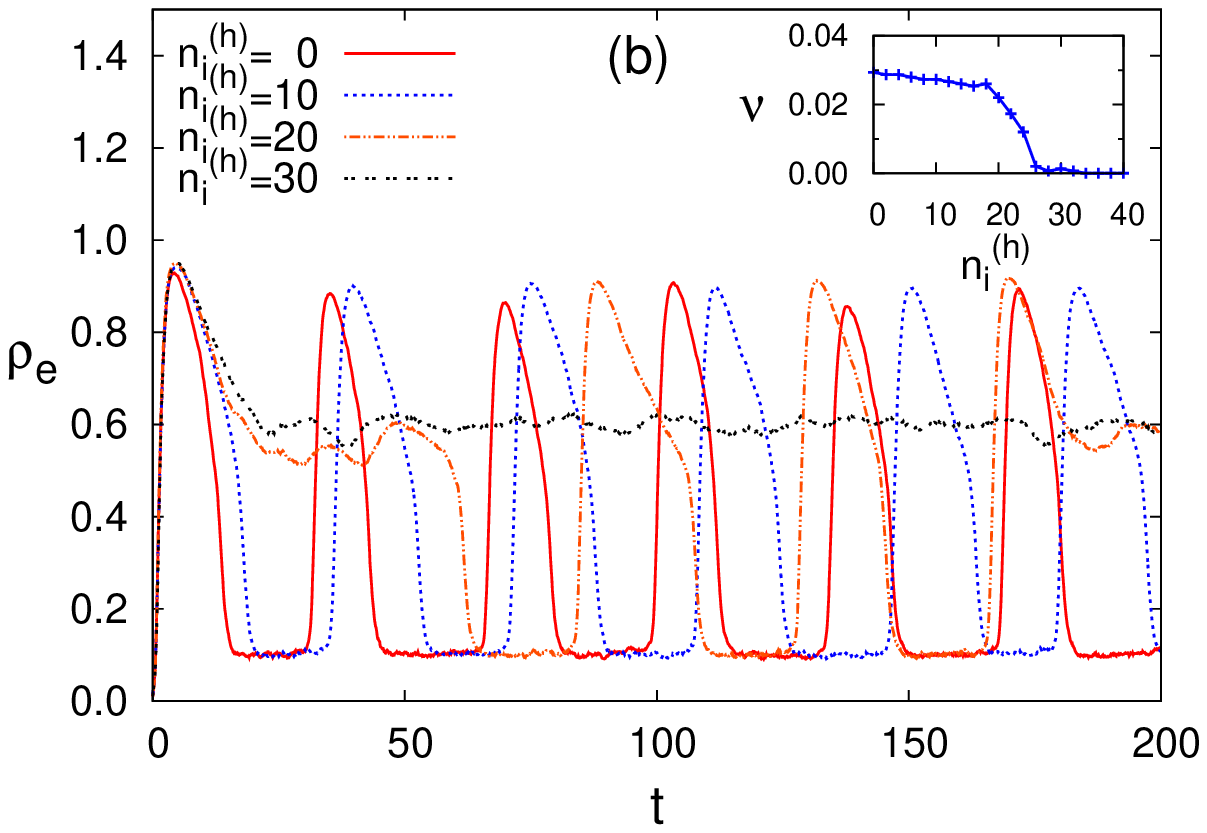}
\ec
\vspace*{-0.4cm}
\caption[]{\small\label{fig:hubremovalunshuffled}
Sustained  oscillations of excitatory activity in  scale-free neuronal networks damaged by targeted attacks.
(a) Removal of $n_e^{(h)}$ excitatory hubs.
(b) Removal of $n_i^{(h)}$ inhibitory hubs.
Insets: Frequency $\nu$ of the sustained oscillations versus the number of removed excitatory or inhibitory hubs. Parameters:
$N= 10000$, $g_i=0.2$, $\gamma=2.5$, $K_{ab}=75$, $F=0.1$,
$J_{i}=-3.5$,
$\Omega=10$,
$\alpha=0.1$,
$\Delta t=0.1$.
}
\end{figure}
Now we study the role of hubs in collective dynamics of neuronal networks with scale-free structure. Using the same network with degree exponent $\gamma=2.5$ as above, we attack excitatory or inhibitory hubs and study sustained network oscillations. In the static model
described in Sec.~\ref{sec:structure}, if a neuron has a small index $j$ then with a high probability
it will have many connections, i.e., it will be a hub. Thus, removal of neurons one by one, starting from $j=1$, corresponds to targeted attacks on hubs.
Fig.~\ref{fig:hubremovalunshuffled} represents impact of these attacks on frequency of sustained network oscillations. One can see that removal of about 120 excitatory hubs or about only 30 inhibitory suppress completely sustained network oscillations. The observed vulnerability of oscillations to attacks is similar to attack vulnerability found in Ref.~\cite{albert00} in other real scale-free networks, such as the
World Wide Web (www), Internet, social networks or a cell. The common properties of these networks, including neuronal networks in the brain, is that their structure is characterized by a fat-tailed degree distribution with divergent second moment.

\section{Conclusion}

In the present paper, we have studied the role of scale-free structure and the impact of noise and damage on dynamics of neuronal networks.
We have developed an analytical approach and performed simulations of dynamics of
a cortical model with stochastic neurons and scale-free organization.
We have found that, in scale-free networks with divergent second moment of degree distribution (networks with degree exponent $2<\gamma\leq 3$), such as functional networks in the human brain),
a very small noise level can stimulate correlated activity of a finite fraction of neurons. The critical level of noise stimulating the activity decreases with increasing
the network size and vanishes in the thermodynamic limit.
The threshold level of noise, that is necessary to stimulate sustained network oscillations,
is also strongly reduced in this kind of scale-free networks. Networks with a finite second moment of degree distribution (scale-free networks with $\gamma >3$)
need a higher noise level to stimulate correlated neuronal activity and this level is finite in the thermodynamic limit.
Furthermore, we have studied impact of random damage on sustained network oscillations in scale-free networks with $2<\gamma\leq 3$.
We demonstrated that, in these networks,
network oscillations are tolerant to random damage in a broad range of the number of randomly removed excitatory or inhibitory neurons. With increasing the number of removed neurons,
the frequency of network oscillations gradually decreases similar to the slowing-down of the alpha rhythm in Alzheimer's disease. However, the networks are vulnerable to targeted attacks on hubs. A targeted attack on a few hubs can impair sustained neuronal oscillations in contrast to the case of random damage. Interestingly, the scale-free networks are more tolerant to removal of excitatory neurons than to removal of inhibitory neurons.
Impact of damage is related to the fact that the removal of neurons decreases the network’s connectivity and, in turn, attenuates synchronization between neurons.

\section{Acknowledgements}
This work was partially supported by the PTDC projects
SAU-NEU/103904/2008, FIS/108476/2008, MAT/114515/2009, and
PEst-C/CTM/LA0025/2011.
D.H. was supported by FCT grant SFRH/BPD/64509/2009.
We also thank Kyoung Eun Lee and Marinho Lopez for valuable discussions.

\appendix

\section{Derivation of rate equations}
\label{derivation}

Here we derive the rate equations (\ref{eq:eqofmotactivit}) and (\ref{eq:EOMactivit})
for the cortical model with stochastic neurons and scale-free structure (the static model in Sec.~\ref{sec:structure}). First, we find a rate equation for the probability $\rho_a(\ell,t)$ that at time $t$ neuron $\ell$ in excitatory or inhibitory population, i.e., $a=e$ or $i$, respectively, is active. For this purpose, we use the method of generating functions from paper \cite{lee04} and generalize it to case of the cortical model with two neuronal populations in Sec.~\ref{sec:structure}. The probability that presynaptic neurons $\ell$ in population $a$ is connected to postsynaptic neuron $j$ in population $b$ is $p_{a,\ell;b,j}$, Eq.~(\ref{eq:palbj}). 
For neuron $j$ in population $b$, we introduce a generating function,
\begin{eqnarray}
&& G_{ab}(j,x,y)=\prod_{\ell=1}^{N_a}\Big\{\exp{(-p_{a,\ell;b,j})} \,\,+
\nonumber\\
&& [x \rho_a(\ell,t)+y(1-\rho_a(\ell,t))](1-\exp{(-p_{a,\ell;b,j})}) \Big\}\approx
\nonumber\\
&&  \exp\Big\{ {-}\!\! \sum_{\ell=1}^{N_a}p_{a,\ell;b,j} {+}\sum_{\ell=1}^{N_a}[x \rho_a(\ell,t){+}y(1{-}\rho_a(\ell,t))]p_{a,\ell;b,j} \Big\},
\nonumber\\
\label{GF-1}
\end{eqnarray}
where we used the fact that $p_{a,\ell;b,j}\ll 1$ at $N \gg 1$. One can show that the probability that neuron $j$ in population $b$ has $n$ active presynaptic neurons in population $a$ at time $t$ is equal to
\begin{equation}
\frac{1}{n!}\frac{\partial^n}{\partial x^n}G_{ab}(j,x,y)\Big|_{x=0,y=1}.
\label{GF-2}
\end{equation}
The condition $y=1$ means that other presynaptic neurons are inactive. Using Eqs.~(\ref{eq:Cabj}) and (\ref{eq:rhotildeb}) for $C_{ab}(j)$ and the weighted activity $\tilde{\rho}_{a}(t)$, we obtain that the probability (\ref{GF-2})  is equal to $P_{n}(\tilde{\rho}_{a}(t)C_{ab}(j))$. Note that this probability depends on the weighted activity $\tilde{\rho}_{a}$, Eq.~(\ref{eq:rhotildeb}), of presynaptic neurons rather than the activity $\rho_{a}$, Eq.~(\ref{eq:rhoa}), that determines the fraction of active neurons in population $a$. This is in contrast to the cortical model on sparsely connected network \cite{goltsev10} where the probability depends on $\rho_{a}$.

Using the probability (\ref{GF-2}), we can find the probability that neuron $j$ in population $b$ has $n$ active presynaptic excitatory and $m$ active presynaptic inhibitory neurons. It is equal to $P_{n}(\tilde{\rho}_{e}(t)C_{eb}(j))P_{m}(\tilde{\rho}_{i}(t)C_{ib}(j))$. This probability allows us to calculate the probability $\Psi_{b,j}(\tilde{\rho}_e, \tilde{\rho}_i)$ that an input at neuron $j$ in population $b$ at time $t$ is at least the threshold $\Omega$, see Eq.~(\ref{eq:psibj}). 

Based on the dynamical rules in Sec.~(\ref{sec:rules}) and the method of Ref. \cite{goltsev10}, we find a change of the probability $\rho_b(j,t)$ during the integration time $\tau$,
\begin{eqnarray}
&&\rho_b(j,t+\tau)-\rho_b(j,t)= \tau f_b [1-\rho_b(j,t)] + \nonumber\\
&&\tau \mu_b [1{-}\rho_b(j,t)] \Psi_{b,j}(\tilde{\rho}_e, \tilde{\rho}_i) {-}\tau \mu_b \rho_b(j,t) [1{-}\Psi_{b,j}(\tilde{\rho}_e, \tilde{\rho}_i)]. \nonumber\\
\label{GF-3}
\end{eqnarray}
The first two terms on the right-hand side describe activation of neuron $j$ by noise and input from presynaptic neurons with the probabilities $\tau f_b$ and $\tau \mu_b $, respectively. The last term describes decay of $\rho_b(j,t)$ with the probability $\tau \mu_b $, if input is smaller than the threshold $\Omega$. Assuming that characteristic time scales of the collective dynamics are much larger than the integration time $\tau$, we obtain a rate equation,
\begin{equation}
\dot{\rho_b}(j,t)=f_b -\nu_b \rho_b(j,t)+ \mu_b \Psi_{b,j}(\tilde{\rho}_e, \tilde{\rho}_i).
\label{GF-4}
\end{equation}
where $\nu_b=f_b + \mu_b$. Multiplying this equation by the weight $w_b(j)$ or $1/N_b$ and summing over $j$, we obtain Eqs.~(\ref{eq:eqofmotactivit}) and (\ref{eq:EOMactivit}), respectively.

\section{Absence of an activation threshold in the static model}
\label{sec:analytcalcs}

\begin{figure}[h]
\bc
\includegraphics[height=6cm, angle=0]{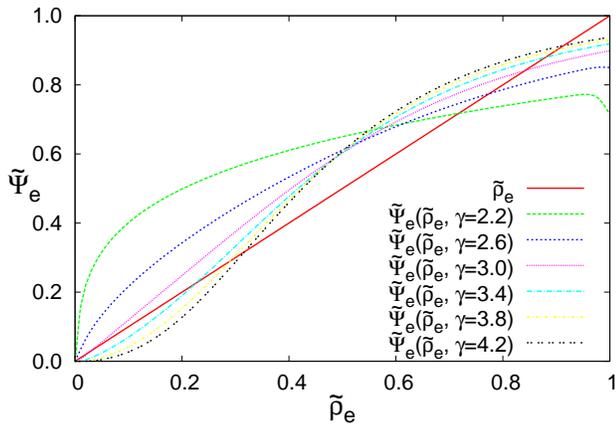}
\ec
\vspace*{-0.4cm}
\caption[]{\small\label{fig:detactivationprob}
Function $\tilde{\Psi}_e(\tilde{\rho}_e, \tilde{\rho}_i)$ defined by Eq.~(\ref{eq:psitildeb})
versus $\tilde{\rho}_e=\tilde{\rho}_i$
at different values of the degree exponent $\gamma$.
$N=10000$, $g_i=0.2$,
$K_{ab}=100$,
$J_{i}=-2$,
$\Omega=18$.
}
\end{figure}

In the steady state, the rate equations~(\ref{eq:eqofmotactivit}) take a form,
\begin{align}
\label{eq:steady}
& \tilde{\rho}_e =F_e +(1-F_e)\tilde{\Psi}_e(\tilde{\rho}_e, \tilde{\rho}_i), \nonumber \\
& \tilde{\rho}_i =F_i +(1-F_i)\tilde{\Psi}_i(\tilde{\rho}_e, \tilde{\rho}_i),
\end{align}
where $F_e$ and $F_i$ are defined by Eq.~(\ref{F}). Solving non-linear Eqs.~(\ref{eq:steady}), one finds a dependence of weighted neuronal activities  $\tilde{\rho}_e$ and $\tilde{\rho}_i$ on the noise level $F$, where $F \equiv F_e=F_i$.
From Eq.~(\ref{eq:psitildeb}), we find that in the case $2<\gamma \leq 3$, the functions $\tilde{\Psi}_e(\tilde{\rho}_e, \tilde{\rho}_i)$  and $\tilde{\Psi}_i(\tilde{\rho}_e, \tilde{\rho}_i)$ have a singular asymptotic behavior at small activity $\tilde{\rho}_e,\tilde{\rho}_i \ll 1$,
\beq
\tilde{\Psi}_a(\tilde{\rho}_e, \tilde{\rho}_i)\approx A \,{\tilde{\rho}}_{e}^{\gamma - 2} \; .
\label{asympt-psi}
\eeq
Therefore, $d \tilde{\Psi}_a / d \tilde{\rho}_e \rightarrow \infty$ at $\tilde{\rho}_e \rightarrow 0$. One can see this singular behavior in Fig.~\ref{fig:detactivationprob} that displays the function $\tilde{\Psi}_e(\tilde{\rho}_e, \tilde{\rho}_i)$
versus $\tilde{\rho}_e$ at $\tilde{\rho}_e=\tilde{\rho}_i$ at different degree exponents $\gamma$.
Substituting Eq.~(\ref{asympt-psi}) into Eqs.~(\ref{eq:steady}), one finds that, in the case $2<\gamma \leq 3$, the only solution of Eqs.~(\ref{eq:steady}) is a solution with $\tilde{\rho}_e$ and $\tilde{\rho}_i$ of order $O(1)$ at any non-zero $F$. However, at $\gamma > 3$, the function $\tilde{\Psi}_e(\tilde{\rho}_e, \tilde{\rho}_i)$ does not have this kind of singularity and $d \tilde{\Psi}_a / d \tilde{\rho}_e \rightarrow 0$ at $\tilde{\rho}_e \rightarrow 0$, see Fig.~\ref{fig:detactivationprob}. In this case, Eqs.~(\ref{eq:steady}) have a stable solution $\tilde{\rho}_e\approx\tilde{\rho}_i\approx F$ at $F \ll 1$.

\section{Activation rate of neurons stimulated by synaptic noise}
\label{sec:approxestim}

In the cortex, the observed mean rate $ \omega_{\operatorname{sn}}^{(0)}$ of spontaneous release of neurotransmitters in a synapsis is in the range
$ [0.01\,\operatorname{Hz}, \, 4\, \operatorname{Hz}]$. This is the source of synaptic noise. If an excitatory or inhibitory neuron has $q$ presynaptic connections, then the rate of random inputs is $\omega_{\operatorname{sn}}=q\,\omega_{\operatorname{sn}}^{(0)}$. If there are no correlations between spontaneous release of neurotransmitters in the synapses and the process is described by a Poisson process, we find that the probability that during the integration time $\tau$ a neuron receives $n$ random inputs is   $P_n(\tau \, \omega_{\operatorname{sn}})$ where $P_n(x)$ is the Poisson distribution function.
The probability that a neuron receives at least $\Omega$ spikes during time $\tau$ is
\beq
P_{\operatorname{act}}\equiv\sum_{n\geq \Omega} P_n(\tau \, \omega_{\operatorname{sn}}).
\label{prob}
\eeq
For simplicity, in these calculations, we neglected a spontaneous release of neurotransmitters in synaptic connection with inhibitory presynaptic neurons and took into account only connections with excitatory neurons. If the mean number of presynaptic connections is $q\approx 1000$,
the integration time $\tau$ is 10ms,
the threshold $\Omega=30$, and the rate $\omega_{\operatorname{sn}}^{(0)}$ in the interval 1-4 Hz, then we obtain that the probability $P_{\operatorname{act}}$ varies from $2\times 10^{-7}$ to 0.96.
In turn, the probability $P_{\operatorname{act}}$ is related with a rate $f$ of the activation process, $P_{\operatorname{act}}=f\tau$. This relationship determines the rate $f$ of activation of neurons by synaptic noise.

\end{document}